\documentclass[twocolumn,showpacs,preprintnumbers,amsmath,amssymb,floatfix]{revtex4}

\usepackage{epsfig}
\usepackage{graphicx}
\usepackage{dcolumn}
\usepackage{bm}

\usepackage{amsmath}
\usepackage{amssymb}
\usepackage{color}
\usepackage{sistyle}

\begin{document}

\preprint{APS/123-QED}
\title{Atomic-like spin noise in solid state demonstrated with manganese in cadmium telluride}

\author{S. Cronenberger, D. Scalbert}
\email{Denis.Scalbert@univ-montp2.fr}
\affiliation{Laboratoire
Charles Coulomb UMR 5221 CNRS/UM2, Universit\'{e} Montpellier, Place
Eugene Bataillon, 34095
Montpellier Cedex 05, France}%
\author{D. Ferrand, H. Boukari, J. Cibert}
\affiliation{Univ. Grenoble Alpes, F-38000 Grenoble, France}
\affiliation{CNRS, Institut NEEL, F-38000 Grenoble, France}

\date{\today}

\maketitle

\textbf{Spin noise spectroscopy is an optical technique which can
probe spin resonances non-perturbatively. First applied to atomic
vapors, it revealed detailed information about nuclear magnetism and
the hyperfine interaction. In solids, this approach has been limited
to carriers in semiconductor heterostructures. Here we show that
atomic-like spin fluctuations of Mn ions diluted in CdTe (bulk and
quantum wells), can be detected through the Kerr rotation associated
to excitonic transitions. Zeeman transitions within and between
hyperfine multiplets are clearly observed in zero and small magnetic
fields and reveal the local symmetry due to crystal field and
strain. The linewidths of these resonances are close to the dipolar
limit. The sensitivity is high enough to open the way towards the
detection of a few spins in systems where the decoherence due to
nuclear spins can be suppressed by isotopic enrichment, and towards
spin resonance microscopy with important applications in biology and
materials science.}
\medskip

Coupling polarized light to atomic spin ensembles non resonantly,
the atom-light
interface\cite{Aleksandrov1981,crookerspectroscopy2004}, is a
powerful approach to manipulate collective spin states, and to
generate spin squeezed \cite{Kuzmich1998,Geremia2004} and entangled
atomic states \cite{Julsgaard2001}. Atomic ensembles, with which
these methods have been developed, are hardly scalable. Scalability
is more straightforwardly achieved in solids, however such an
approach has been limited to carriers in semiconductor
heterostructures
\cite{crookerspin2010,Yan2012,hornspin2013,zapasskiioptical2013,dahbashioptical2014,
poltavtsevspin2014,yangtwo-colour2014}.

Here, we introduce the atom-exciton-light interface in solids, to transfer the atomic spin noise to light polarization in a system where the atomic spin is not directly coupled to light: in the case of Mn in CdTe which we use as a testbed, the exciton mediates the coupling between atoms and
light. This interface is based on the sp-d exchange interaction,
which couples the manganese ions to the carriers. Thus atomic spin
fluctuations produce tiny splittings of the  excitonic transitions,
ultimately detected by the induced circular birefringence (Kerr
rotation). We show that the spin noise spectra keep their atomic
nature despite the strong hybridization between the orbitals of
manganese ion and the crystal. This opens new possibilities by
transposing in solids the method of quantum physics, based on the
atom-light interface.

\vspace{0.5 cm} \begin{large} \textbf{Results} \end{large}
\vspace{0.5 cm}

\textbf{Spin Hamiltonian and the spin-noise spectra.} Electron spin
resonances can be described by a spin Hamiltonian, in which the
allowed terms are dictated by symmetry considerations. For Mn in a
tetrahedral crystal field it takes the form \cite{AbragamBleaney}

\begin{equation}\label{Hspin}
\begin{split}
H& =A \mathbf{I}\cdot \mathbf{S}+g\mu_\text{B}\mathbf{B}\cdot \mathbf{S}\\
& +\frac{a}{6}[S^{4}_\text{x}+S^{4}_\text{y}+S^{4}_\text{z}-\frac{1}{5}S(S+1)(3S^2+3S-1)]
\end{split}
\end{equation}

Here \textbf{I} and \textbf{S} are the Mn nuclear and electronic
spins respectively,  $\mu_B$ is the  Bohr magneton and \textbf{B} is
the magnetic field. The Hamiltonian is composed of the hyperfine term, with $A=-170.5 \text{ MHz}$, the Zeeman term, with $g=2$ the Mn $g$-factor, and the cubic crystal field term, with $a=89.5 \text{ MHz}$ \cite{Causa1980}.

\begin{figure}
\includegraphics[width=8.8cm] {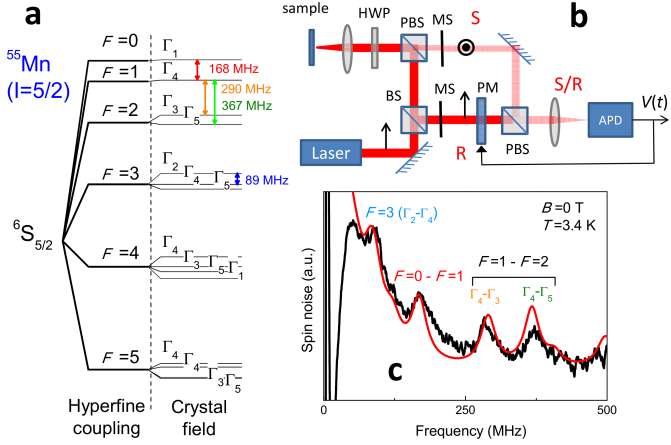}
\caption{\textbf{Spin noise of $^{55}$Mn diluted in bulk CdTe.} \textbf{a,} Energy levels of $^{55}$Mn split by the hyperfine coupling and the cubic crystal field. Arrows point the transitions observed in c. \textbf{b}, Experimental set-up. A vertically polarized laser beam is focused on the sample. Mn spin fluctuations impart Kerr rotation noise on the reflected probe which is split into two mutually orthogonal polarizations (beam S and R) by a polarizing beamsplitter (PBS). The beam S carries the spin fluctuations while beam R carries only the intensity fluctuations. Both beams are sent alternatively on the avalanche diode (APD) using mechanical shutters (MC), and kept at the same intensity with a feedback loop on the phase modulator (PM). \textbf{c}, Spin noise spectra in zero magnetic field, as measured (black line) and calculated (red line) with only one parameter $\gamma=100$~MHz. The dip in the experimental spectrum below 50~MHz comes from subtraction of a zero frequency line not accounted for by theory (see text).}
\label{scheme}
\end{figure}

The $^6S_{5/2}$ Mn ground state is split by the hyperfine coupling
in six F-levels (Fig.~\ref{scheme}a). Without crystal field, the
F-levels are split by a magnetic field into $2F+1$ Zeeman levels,
with a common $g$-factor $g_F=1$ because $S=I$. The crystal field
further splits and mixes the hyperfine levels, which considerably
increases the number of allowed transitions \cite{AbragamBleaney}.

According to the fluctuation-dissipation theorem the spectrum of the spin fluctuations for a single Mn spin in the direction of the unit vector
$\mathbf{\hat{\alpha}}$, at thermal equilibrium, is related to the susceptibility and is given by \cite{Landau}

\begin{equation}
\label{NoiseDSP}
( \mathbf{S}\cdot \mathbf{\hat{\alpha}} )^2_\omega= \sum_{n,m}
(\rho_{n}+\rho_{m}) | \langle n|\mathbf{S}\cdot
\mathbf{\hat{\alpha}}|m\rangle |^2
\frac{\gamma}{(\omega+\omega_{nm})^2+\gamma^2}
\end{equation}

where the double summation is over the $(2S+1)\times(2I+1)=36$
eigenstates of the spin hamiltonian, $\rho_n$ is the occupation
factor at thermal equilibrium of level $n$ with eigenfrequency $\omega_n$ and
$\omega_{nm}=\omega_m-\omega_n$. For simplicity, we assume
Lorentzian lines with a broadening parameter $\gamma$, common
to all transitions, and written as $\gamma_\text{T}$ for the transverse configuration ($\textbf{B} \bot
\mathbf{\hat{\alpha}}$) and $\gamma_\text{L}$ for the longitudinal geometry ($\textbf{B}$ // $\mathbf{\hat{\alpha}}$).

In the transverse configuration, Zeeman transitions within each
hyperfine level contribute to a single line at $g_F=1$, and
inter-hyperfine transitions between adjacent $F$-levels contribute
to higher frequency lines. The sum rules derived in Supplementary
Note 1 shows that half of the integrated spin noise is concentrated
in the $g_F=1$ line, the other half being shared between all
inter-hyperfine transitions. In the longitudinal case,
$\mathbf{S}\cdot \mathbf{\hat{\alpha}}$ is diagonal within each
hyperfine level and the corresponding noise signal appears around
zero frequency. The crystal field brings in some complexity but does
not affect significantly the intensity share between intra- and
inter- hyperfine transitions.

\textbf{Samples and experiment.} All the results are obtained from
Cd$_{0.999}$Mn$_{0.001}$Te samples: a bulk crystal cleaved along a
(110) plane and quantum wells (QW) grown either on the (001) plane
of CdTe or Cd$_{96}$Zn$_{0.04}$Te substrates. The spin fluctuations
are probed along the laser beam, perpendicular to the sample
surface. More details about the samples and the experimental set-up
(Fig. 1b) can be found in  section Methods.
\medskip

\begin{figure}
\includegraphics[width=8.8cm]{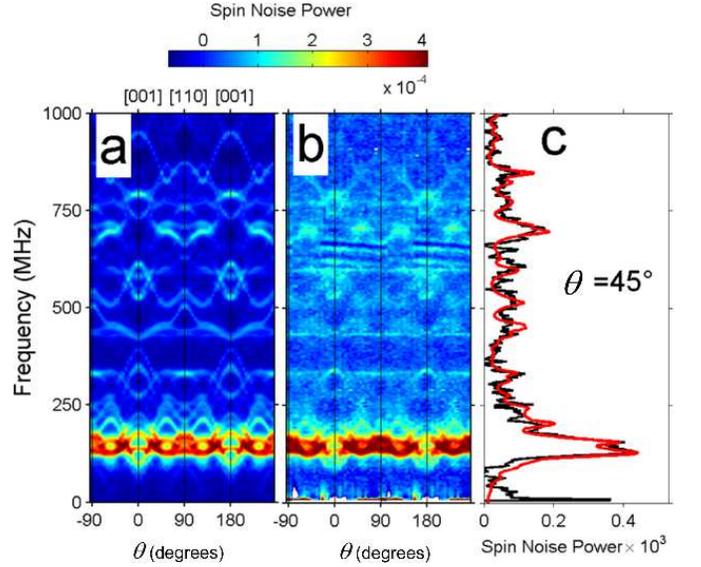}
\caption{\textbf{Angular resolved spin noise of $^{55}$Mn diluted in bulk CdTe at $B_T$=10.5~mT. a}, Contour plot of the spectra calculated with Eq. (\ref{NoiseDSP}) versus $\theta$, the angle between magnetic field and the [001] direction contained in the (110) plane. \textbf{b}, Experimental contour plot ($T=4.8$~K). \textbf{c}, Experimental spectrum at $\theta=45^{\circ}$ (black line) and best fit (red line) with $\gamma_T=$50~MHz.}
\label{FigCMT300}
\end{figure}

\textbf{Bulk sample.} The spin noise spectrum of Mn diluted in the
bulk crystal is reported in Fig.~\ref{scheme}c. The four strongest
resonances are identified as intra-hyperfine, and inter-hyperfine
transitions between states of the Mn split by the cubic crystal
field (Fig.~1a). The data are well reproduced by the model with as
only fitting parameter the HWHM in omega units $\gamma=100$~MHz.
Spin noise spectra have also been measured with a magnetic field
applied in all directions of the (110) plane defined by the angle
$\theta$ (by steps of $5^{\circ}$). As expected, we obtain a
twofold symmetry in the frequency map. We then average the noise
spectra measured at $\theta$ and $\theta+180^{\circ}$ in order to
better resolve the weak structures, which spread over the 1~GHz
width (Fig.~\ref{FigCMT300}b, c). All the dominant features
predicted by equation~(\ref{NoiseDSP}) can be recognized in the
experimental spectra and many details are perfectly reproduced by
the simulation with the linewidth $\gamma_\text{T}$ being the only fitting
parameter. A line centered at zero frequency, not predicted by the
theory, has been subtracted from the experimental spectra to allow
the comparison with theory: the original spectra are given in
Supplementary Fig.~1.

\medskip
\textbf{Quantum wells.} The spin noise spectra are quite sensitive
to small lattice distortions. The presence of a lattice mismatch
between the QW and substrate imposes adding a spin anisotropy term
$D_0 S_\text{z}^2$ to the spin hamiltonian. This term is clearly observed
in the QW grown on Cd$_{96}$Zn$_{0.04}$Te, with the expected value
$D_0=+473\text{ neV}$ obtained using the known Mn spin-lattice
coefficients \cite{Causa1980}. However, the fourfold cubic symmetry
is broken. Adding an in-plane anisotropy term partially reproduces
the experimental spectra (see Supplementary Fig.~2).

\begin{figure}
\includegraphics[width=8.7cm] {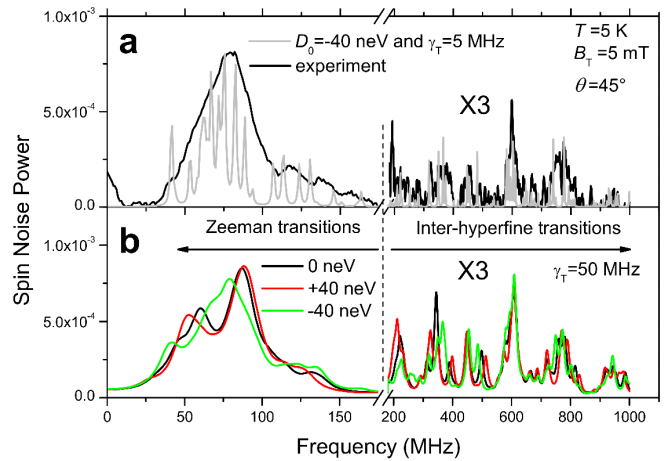}
\caption{\textbf{Adjustment of the biaxial strain to fit the experimental spin noise spectra of $^{55}$Mn in the QW grown on CdTe.} $\textbf{a}$, Experimental (black line) and calculated (grey line) spin noise spectra. As shown by the calculation with a small broadening $\gamma_\text{T}=5$~MHz, many unresolved individual lines, which belong either to the Zeeman transitions (left) or to inter-hyperfine transitions (right), are unresolved in the experimental spectrum. $\textbf{b}$, Calculated noise spectra for different values of the strain parameter $D_0$: the best agreement with the experiment is obtained for $D_0=-40 \text{ neV}$ and $\gamma_\text{T}=50$~MHz (green curve).}
\label{fit_M3129_5mT}
\end{figure}

\begin{figure}
\includegraphics[trim=0cm 0 0cm 0, clip=true, width=8.8cm] {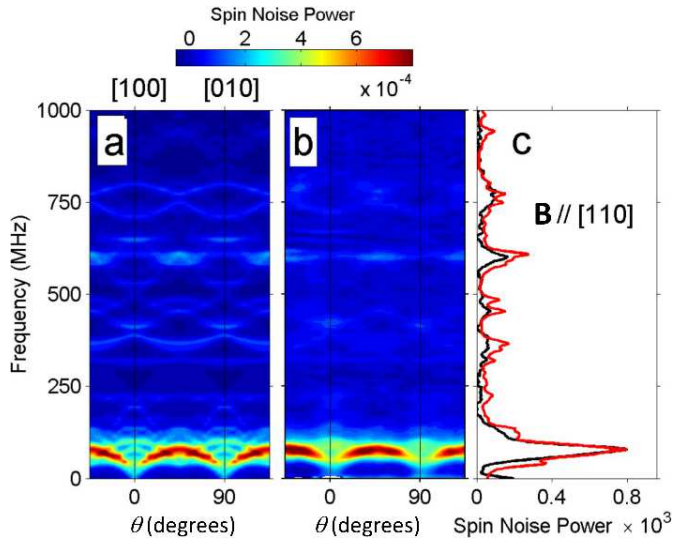}
\caption{\textbf{Angular resolved spin noise of $^{55}$Mn in the QW grown on CdTe at $B_T$=5~mT. a}, Contour plot of spin noise spectra calculated for $\gamma_T=50$~MHz, and $D_0=-40$~neV. \textbf{b}, Experimental contour plot ($T=5$~K). \textbf{c}, Experimental (black line) and  calculated spectra (red line) at $\theta=45^{\circ}$ (B//[110]).}
\label{M3219_5mT}
\end{figure}

\begin{figure}
\includegraphics[trim=0cm 0 0cm 0,clip=true,width=8.8cm] {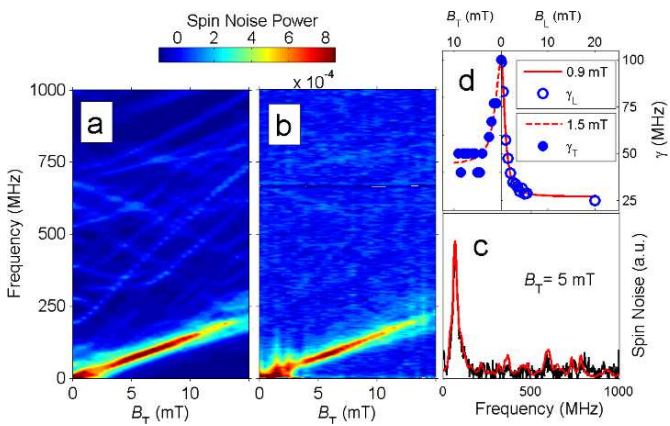}
\caption{\textbf{Spin noise spectra of $^{55}$Mn in the QW grown on
CdTe versus magnetic field intensity at $\theta=30^{\circ}$. a},
Contour plot of spin noise spectra calculated for $\gamma_T=50$~MHz,
and $D_0=-40$~neV. \textbf{b}, Experimental contour plot ($T=6$~K).
\textbf{c}, Fitting Eq. (\ref{NoiseDSP}) to the experimental
spectrum determines $\gamma$. \textbf{d}, Obtained values of
$\gamma_T$ in transverse field and $\gamma_L$ longitudinal fields
(spectra in longitudinal fields given in Supplementary Fig.~4). Red
curves are fits of $\gamma_L$ and $\gamma_T$ to lorentzian
(HWHM=$0.9\text{ mT}$ and $1.5\text{ mT}$, respectively).}
\label{M3129_m15deg}
\end{figure}

Including a $D_0 S_\text{z}^2$ term also in the quasi lattice-matched QW grown on CdTe is
sufficient to obtain a relatively good agreement between experiment
and theory (see Fig.~(\ref{fit_M3129_5mT})). The anisotropy
coefficient $D_0=-40\text{ neV}$ is significantly larger than the
calculated value $D_0=-5\text{ neV}$. The expected
fourfold cubic symmetry for magnetic fields applied in the (001) plane of the QW  is observed (Fig.~(\ref{M3219_5mT})). Figure
(\ref{fit_M3129_5mT}a) also reveals that the line centered at
$75\text{ MHz}$ consists in many unresolved individual lines.
Figure~(\ref{M3129_m15deg}) shows that this bunch of lines shifts
almost linearly with the magnetic field, and can be assigned to the
Zeeman transitions with $g_F=1$. These lines are spread in frequency
both by the crystal field, and by the quadratic Zeeman effect which
arises because of the gradual decoupling of electronic and nuclear
spins. The quadratic Zeeman effect becomes more visible above
$10\text{ mT}$ where the $g_F=1$ line progressively broadens due to its splitting in many unresolved individual lines.
Barely visible transitions at higher frequencies correspond to
inter-hyperfine transitions $\Delta F=1$. We point out that all the
peaks of the unconventional spectra revealed by our low-field
measurements, must evolve and contribute at high field to the
well-known structure of the Mn spin resonance spectra consisting of
6 equally spaced lines.

\textbf{Linewidths.} Our results give new insights in the Mn spin
relaxation mechanisms at low magnetic field, a regime usually not
accessible by conventional spin resonance techniques. In the
transverse configuration, although the presence of many unresolved
lines complicates the determination of the broadening parameter
$\gamma_T$, it can be estimated by fitting the experimental spectra
with equation~(\ref{NoiseDSP}) (see Fig.~(\ref{M3129_m15deg}c)).
Figure~(\ref{M3129_m15deg}d) shows that $\gamma_\text{T}$ notably decreases
from 100~MHz  at zero field to 40~MHz above a characteristic field
$B_\text{c}\sim 1\text{ ~mT}$.

A potential source of broadening in this range of Mn composition $x$, is the
dipolar interactions among the ensemble of electronic Mn spins
\cite{Abragam}. Here, a Lorentzian shape is expected, with
the wings of the line formed by spins with a first neighbor at short
distance, and the center due to spins with no neighbor in a volume
$\sim1/xN_0$, where $N_0=4/a^3$ is the density of cation sites in
CdTe, with a cubic lattice parameter $a$=0.648~nm. Adapting the
calculation of moments of Ref.~\onlinecite{Kittel1953} to a Mn spin
with $S=5/2$, and $g=2$, we obtain an effective field $B_\text{eff}\simeq5~g\mu_\text{B}~ N_0x
\sqrt{S(S+1)}~ \mu _0/4 \pi=0.4~\text{mT}$.
Assuming that the effect of hyperfine coupling affects the Land\'{e}
factor and not $B_\text{eff}$, the resulting linewidth with $g=1$ is
$\gamma=g\mu_\text{B} B_\text{eff}/\hbar=35~\text{MHz}$, to be increased by a
factor $10/3$ at zero field \cite{Abragam}.

In a longitudinal field, $\gamma_\text{L}$ is much easier to measure, as
all the intra-hyperfine lines merge at zero frequency. Spectra are
given in Supplementary Fig.~3. We expect the dipolar broadening to
be totally suppressed as soon as the applied field is larger than
the dipolar fields. This is the case in figure~(\ref{M3129_m15deg}d)
where $\gamma_\text{L}$ rapidly decreases; however it saturates at a finite
value, suggesting that a relaxation mechanism, with
$T_{1}=\gamma_\text{L}^{-1}$, remains to be identified.

In both cases, a simple estimation of dipolar broadening
qualitatively explains the behavior of the linewidth upon applying a
magnetic field, and gives reasonable orders of magnitude. However,
there is another, smaller contribution which appears to be quite
independent of the applied field. Estimates of other sources of
broadening are given in Supplementary Note 2. Supplementary
Fig.~4 shows that the linewidth is robust to changes of temperature
and excitation power.

\medskip

\vspace{0.5 cm}
\begin{large}
\textbf{Discussion}
\end{large}
\vspace{0.5 cm}

As an outlook, we emphasize that the present measurements can be
extended to higher magnetic fields by using large bandwidth spin
noise spectroscopy techniques \cite{muller2010,Hubner}. This will
permit to study the Paschen-Back regime, when electronic and nuclear
spins are completely decoupled. In this regime, because of the
hyperfine interaction, the Mn electronic spin precession frequency
depends on the orientation of the nuclear spin relative to the
applied field \cite{cronenbergeroptical2013}. The fluctuation
spectra will therefore directly reveal the nuclear spin populations,
an information which cannot be easily addressed by other
spectroscopic methods.

The sensitivity is high enough to open the way towards the detection
of a few spins in systems where the decoherence due to nuclear spins
can be suppressed by isotopic enrichment
\cite{Balasubramanian,tyryshkin2011,Sleiter2013}, and towards spin
resonance microscopy with important applications in biology and
materials science \cite{blankhigh2003}. Additional coupling of the
spins to a microcavity should enable single spin detection
\cite{dahbashioptical2014,Salis2005,Hu2008}: this will open a
fascinating opportunity to explore quantum jumps of the Larmor
frequency of a single Mn spin (see Supplementary Note 2), and
to realize a high-fidelity readout of its nuclear spin
\cite{plahigh-fidelity2013}.

\vspace{0.5 cm}
\textbf{Methods}

\vspace{0.5 cm}

\begin{small}

\textbf{Quantum well samples.} Three quantum wells (QW) were grown
by molecular beam epitaxy on (100) substrates. A 14~nm wide
CdTe/Cd$_{0.75}$Mg$_{0.25}$Te QW was grown on CdTe as a reference:
no spin noise signal could be detected for this sample without Mn.
Two Cd$_{1-x}$Mn$_x$Te/Cd$_{0.75}$Mg$_{0.25}$Te were grown in the
same conditions as the reference QW: a 14~nm wide QW grown on CdTe
and a 20~nm wide QW grown on Cd$_{0.96}$Zn$_{0.04}$Te. Using
Vegard's law and the lattice parameters of CdTe and ZnTe
\cite{LandoltBorstein} and MnTe \cite{Janik}, one can expect a shear
strain along the growth axis, isotropic in the growth plane, of
\textbf{$2\times10^{-5}$}, in tension, for the Cd$_{1-x}$Mn$_x$Te QW
grown on CdTe and \textbf{$-2\times 10^{-3}$}, in compression, for
the one grown on Cd$_{0.96}$Zn$_{0.04}$Te. The Mn compositions,
x=0.001, have been controlled by magneto-reflectivity.
Photoluminescence revealed a trion line, likely due to residual or
photogenerated holes in the QWs.

\textbf{Experimental setup.} We developed a specific setup adapted
to Mn spin noise spectroscopy, which requires a large bandwidth up
to 1~GHz, and a high sensitivity. To that purpose we used an
avalanche diode with a very low noise equivalent power (typical NEP
0.4~pW/Hz$^{1/2}$), and a short response time of 0.5~ns. The
sensitivity is then maximized by detecting the spin fluctuations in
nearly crossed polarization (average power on the detector is less
than $\sim1\mu$W) , while keeping a relatively high probe power on
the sample (typically $\sim 1$~mW in our experiments)
\cite{glasenappresources2013}. In these conditions, the attenuated
laser is at the shot noise. Subtraction of the photon noise from the
total noise is achieved by alternatively measuring the power spectra
of the signal and reference beams (see Fig. \ref{scheme}a). The
normalized spin noise power in units of the photon noise, and
corrected from the apparatus response function is then given by
$S_\text{spin}(f)=\frac{S_\text{signal}(f)-S_\text{ref}(f)}{S_\text{signal}(f)+S_\text{ref}(f)}$.
The samples are mounted on the cold finger of a helium cryostat, and
placed at the center of two-axis Helmholtz coils. One narrowband
continuous-wave diode laser (from Toptica) is tuned to the excitonic
resonance of the bulk or of the QWs (see Supplementary Fig.~5). With a laser spot size
$\sim5\mu m$ ($\sim 4\times 10^6$ Mn atoms lie within the
spot size in the case of the QW).  The spin noise signal is
continuously digitized at 2~GHz and processed by a
field-programmable gate array (Agilent card U1080A), to obtain the
spin noise power spectrum. Typically, each spectrum requires a few
minutes of signal averaging.

\end{small}

\vspace{0.5 cm}

\textbf{Acknowledgements.}
D.S. and S.C warmly acknowledge stimulating discussions with M. Dyakonov and M. Myara. This work was supported by the French ANR research project SNS (Grant No. 2011-BS04-018 01).

\vspace{0.5 cm}

\textbf{Author contributions.}
D.S. supervised the whole project. D.S. and S.C. conceived the experiment, built the experimental apparatus, and analyzed the data. D.S. wrote the manuscript in close collaboration with all authors, particularly J.C.. H.B grew the samples. D.F. performed the magneto-optical spectroscopy of the quantum wells. All authors contributed to numerous discussions and revised the manuscript.

\vspace{0.5 cm}

\textbf{Competing financial interests.}
 The authors declare no competing financial interests.


\begin{thebibliography}{30}
\bibitem {Aleksandrov1981}  Aleksandrov, E. B. \& Zapassky, V. S. Magnetic resonance in the Faraday-rotation noise spectrum. \emph{Zh. Eksp. Teor. Fiz.} \textbf{81,} 132-138
(1981).
\bibitem{crookerspectroscopy2004}   Crooker, S.~A., Rickel, D.~G., Balatsky, A.~V. \& Smith, D.~L. Spectroscopy of spontaneous spin noise as a probe of spin dynamics and magnetic resonance. \emph{Nature} \textbf{431,} 49-52
(2004).
\bibitem {Kuzmich1998}  Kuzmich, A., Bigelow, N. P. \& Mandel, L. Atomic quantum non-demolition measurements and squeezing. \emph{EPL} \textbf{42,} 481 (1998).
\bibitem{Geremia2004}   Geremia, J., Stockton, J.~K. \& Mabuchi, H. Real-time quantum feedback control of atomic spin-squeezing. \emph{Science} \textbf{304,} 270-273 (2004).
\bibitem{Julsgaard2001} Julsgaard, B., Kozhekin, A. \& Polzik, E.~S. Experimental long-lived entanglement of two macroscopic objects. \emph{Nature} \textbf{413,} 400-403 (2001).
\bibitem{crookerspin2010}   Crooker, S. \emph{et~al.} Spin noise of electrons and holes in self-assembled quantum dots. \emph{Phys. Rev. Lett.} \textbf{104,} 036601 (2010).
\bibitem{Yan2012}   Li, Y. \emph{et~al.} Intrinsic spin fluctuations reveal the dynamical response function of holes coupled to nuclear spin baths in (In,Ga)As quantum dots. \emph{Phys. Rev. Lett.} \textbf{108,} 186603 (2012).
\bibitem{hornspin2013}  Horn, H. \emph{et~al.} Spin noise spectroscopy of donor-bound electrons in ZnO. \emph{Phys. Rev. B }\textbf{87,} 045312 (2013).
\bibitem{zapasskiioptical2013}  Zapasskii, V.~S. \emph{et~al.} Optical spectroscopy of spin noise. \emph{Phys. Rev. Lett.} \textbf{110,} 176601 (2013).
\bibitem{dahbashioptical2014}   Dahbashi, R., Hubner, J., F.,Berski, Pierz, K. \& Oestreich, M. Optical spin noise of a single hole spin localized in an (InGa)As quantum dot. \emph{Phys. Rev. Lett.} \textbf{112,} 156601 (2014).
\bibitem{poltavtsevspin2014}    Poltavtsev, S.~V. \emph{et~al.}, Spin noise spectroscopy of a single quantum well microcavity. \emph{Phys. Rev. B} \textbf{89,} 081304(R) (2014).
\bibitem{yangtwo-colour2014}    Yang, L. \emph{et~al.} Two-colour spin noise spectroscopy and fluctuation correlations reveal homogeneous linewidths within quantum-dot ensembles. \emph{Nat. Commun.} \textbf{5,} 4949 (2014).
\bibitem{Abragam}   Abragam, A. \& Bleaney, B. \emph{Electron Paramagnetic Resonance of Transition Ions}, Clarendon Press - Oxford
(1970).
\bibitem{Causa1980} Causa, M., Tovar, M., Oseroff, S.~B., Calvo, R. \& Giriat, W. Spin-lattice coefficients of Mn$^{2+}$ in II-VI compounds. \emph{Phys. Lett. A }\textbf{77,} 473-475 (1980).
\bibitem{Landau}    Landau, L.~D.\& Lifshitz, E.~M. \emph{Statistical Physics.} Pergamon Press (1980).
\bibitem{AbragamBleaney}   Abragam, A. \emph{Principles of Nuclear Magnetism.} Oxford University Press, New York
(1961).
\bibitem{Kittel1953}    Kittel, C. \& Abrahams, E. Dipolar broadening of magnetic resonance lines in magnetically diluted crystals. \emph{Phys. Rev.} \textbf{90,} 238-239 (1953).
\bibitem{muller2010}    M\"{u}ller, G.~M., R\"{o}mer, M. H\"{u}bner, J. \& Oestreich, M. Gigahertz spin noise spectroscopy in n-doped bulk GaAs. \emph{Phys. Rev. B} \textbf{81,} 121202(R) (2010).
\bibitem{Hubner}    H\"{u}bner, J. \emph{et~al.} Rapid scanning of spin noise with two free running ultrafast oscillators. \emph{Opt. Express }\textbf{21,} 5872-5878 (2013).
\bibitem{cronenbergeroptical2013}   Cronenberger \emph{et~al.} Optical pump-probe detection of manganese hyperfine beats in (Cd,Mn)Te crystals. \emph{Phys. Rev. Lett.} \textbf{110,} 077403 (2013).
\bibitem{Balasubramanian}   Balasubramanian, G. \emph{et~al.} Ultralong spin coherence time in isotopically engineered diamond. \emph{Nature Mater.} \textbf{8,} 383-387 (2009).
\bibitem{tyryshkin2011} Tyryshkin, A.~M. \emph{et~al.} Electron spin coherence exceeding seconds in high-purity silicon. \emph{Nature Mater.} \textbf{11,} 143-147 (2011).
\bibitem{Sleiter2013}   Sleiter, D.~J. \emph{et~al.} Optical pumping of a single electron spin bound to a fluorine donor in a ZnSe nanostructure. \emph{Nano Lett.} \textbf{13,} 116-120 (2013).
\bibitem{blankhigh2003} Blank, A., Dunnam, C.~R., Borbat, P.~P. \& Freed, J.~H. High resolution electron spin resonance microscopy. \emph{J. Magn. Reson.} \textbf{165,} 116-27 (2003).
\bibitem{Salis2005} Salis, G. \& Moser, M. Faraday-rotation spectrum of electron spins in microcavity-embedded GaAs quantum wells. \emph{Phys. Rev. B} \textbf{72,} 115325 (2005).
\bibitem{Hu2008}    Hu, C.~Y., Young, A., O'Brien, J.~L., Munro, W.~J. \& Rarity, J.~G. Giant optical Faraday rotation induced by a single-electron spin in a quantum dot: applications to entangling remote spins via a single photon. \emph{Phys. Rev. B} \textbf{78,} 085307 (2008).
\bibitem{plahigh-fidelity2013}  Pla, J.~J. \emph{et~al.} High-fidelity readout and control of a nuclear spin qubit in silicon. \emph{Nature} \textbf{496,} 334-338 (2013).
\bibitem{LandoltBorstein} Landolt-B\"{o¨}rnstein numerical data and functional relationships in science and technology. Group 3, Crystal and solid state physics. Vol.17, Semiconductors, Springer-Verlag, Berlin (1982).
\bibitem{Janik} Janik, E. \emph{et~al.} Structural properties of cubic MnTe layers grown by MBE. \emph{Thin Solid Films} \textbf{267,} 74-78 (1995).
\bibitem{glasenappresources2013}  Glasenapp, P. \emph{et~al.}   Resources of polarimetric sensitivity in spin noise spectroscopy. \emph{Phys. Rev. B} \textbf{88,} 165314 (2013).





\end{thebibliography}
\end{document}